\documentclass[pra,twocolumn]{revtex4}
\usepackage{amssymb}
\usepackage{amsmath}
\usepackage{graphicx}
\usepackage{dcolumn}
\usepackage[center]{subfigure}
\usepackage{array}
\usepackage{booktabs}
\usepackage{multirow}
\usepackage{color}

\begin{document}

\title{Adiabatic geometric phase in fully nonlinear three-wave mixing}
\author{Yongyao Li$^{1,2}$, Ofir Yesharim$^{2}$, Inbar Hurvitz$^{2}$, Aviv Karnieli$^{2}$, Shenhe Fu$^{3}$, Gil Porat$^{4}$, Ady Arie$^{2}$}
\email{ady@tauex.tau.ac.il}
\affiliation{$^{1}$School of Physics and Optoelectronic Engineering, Foshan University,
Foshan 528000, China \\
$^{2}$ Department of Physical Electronics, School of Electrical Engineering,
Faculty of Engineering, Tel Aviv University, Tel Aviv 69978, Israel.\\
$^{3}$Department of Optoelectronic Engineering, Jinan University, Guangzhou 510632, China\\
$^{4}$Department of Electrical and Computer Engineering, University of Alberta, Edmonton, Alberta T6G 1H9,
Canada}

\begin{abstract}
In a nonlinear three wave mixing process, the interacting waves can accumulate an adiabatic geometric phase (AGP) if the nonlinear coefficient of the crystal is modulated in a proper manner along the nonlinear crystal. This concept was studied so far only for the case in which the pump wave is much stronger than the two other waves, hence can be assumed to be constant. Here we extend this analysis for the fully nonlinear process, in which all three waves can be depleted and we show that the sign and magnitude of the AGP can be controlled by the period, phase and duty cycle of the nonlinear modulation pattern. In this fully nonlinear interaction, all the states of the system can be mapped onto a closed surface. Specifically, we study a process in which the eigenstate of the system follows a circular rotation on the surface. Our analysis reveals that the AGP equals to the difference between the total phase accumulated along the circular trajectory and that along its vertical projection, which is universal for the undepleted (linear) and depleted (nonlinear) cases. Moreover, the analysis reveals that the AGPs in the processes of sum-frequency generation and difference-frequency generation have opposite chirality. Finally, we utilize the AGP in the fully nonlinear case for splitting the beam into different diffraction orders in the far field.
\end{abstract}

\maketitle

\section{Introduction}
\label{sec:intro}
When an eigenstate of a quantum system follows a closed path by slowly varying the parameters of its Hamiltonian, an adiabatic geometric phase (AGP) is acquired, in addition to the dynamical phase \cite{Pancharatnam,Berry}. In past decades, the creation and the application of AGPs in various systems have an been active topic of research in quantum physics \cite{Biaowu, Leek, Yale, YCho}, condensed matter physics \cite{NQ, Yunbo, Mengxiao} and optics \cite{Hariharan, Slussarenko, LAA}. AGP can also be accumulated in a nonlinear optical process, either by using circular polarized light on metasurfaces \cite{Tymchenko2015,GLi2015,Devlin,Haowen}, or by relying on the spectral degree of freedom of the interacting waves, by varying the nonlinear modulation pattern along the nonlinear crystal \cite{Wang2017,Aviv}. Furthermore, wavefront shaping and non-reciprocal transmission were realized experimentally by creating the AGP through circular rotation of the quasi-phase matching (QPM) parameters in the $\chi^{(2)}$ nonlinear process using sum-frequency generation (SFG) \cite{Aviv2}. The analysis of AGP in nonlinear processes was based so far on the undepleted pump approximation: the weak idler and signal are coupled by a strong pump. In this case, the photon fluxes of two waves (idler and signal) conserve to a total norm, and they also construct two eigenstates that satisfy the superposition principle. Hence, such a nonlinear process can be analyzed as a linear system and can be described by spin-1/2 dynamics \cite{Aviv,Suchowski,SPA}. In this case, the AGP can be well predicted via the available knowledge on the dynamics of two level systems. However, when the intensities of the signal and idler are comparable to that of the pump, the undepleted pump approximation becomes invalid since the superposition principle and the conservation of norm cannot be assumed anymore. Adiabatic frequency conversion in the fully nonlinear case was studied in recent years \cite{Porat,Yaakobi,Phillips1}, a criterion for adiabaticity was defined \cite{Gil} and fully nonlinear adiabatic processes were demonstrated experimentally, e.g. quasi-phase matched second harmonic generation \cite{Leshem,Dahan}, birefringently phase matched second harmonic generation \cite{Rozenberg} and parametric amplification \cite{Markov}. Nevertheless, how to control and predict the AGP in the fully nonlinear case remains an open question \cite{Alber}.

In this paper, we aim to solve this question in a pump-depleted $\chi^{(2)}$ process by adiabatically varying the QPM parameters. A novel geometric analysis is provided for a simple and effective prediction. Furthermore, we verify the analysis not only by the SFG process but also for the difference-frequency generation (DFG) process. The rest of the paper is structured as follows. A geometric description of the dynamics of the pump-depleted three-wave mixing (TWM) process under a circular rotation of three QPM parameters is developed in section 2. An optical interferometric scheme for measuring the AGP and a theoretical analysis of how to calculate the AGP are illustrated by the pump-depleted SFG process in sections 3 and 4, respectively. AGP for the pump-depleted DFG process is discussed in section 5.  In section 6, we discuss a potential application of utilizing AGP for all-optical beam shaping in the fully nonlinear regime. The main results of the paper are summarized in section 7.

\section{Geometric representation of three wave mixing}
Following the derivation of Luther et al  \cite{Luther}, the dynamical dimensionless coupled wave equations for TWM process are governed by

\begin{eqnarray}
&&{dq_{1}/d\tau}=i\Delta\Gamma q_{1}-igq^{\ast}_{2}q_{3}, \label{TTbasieq1}\\
&&{dq_{2}/d\tau}=i\Delta\Gamma q_{2}-igq^{\ast}_{1}q_{3}, \label{TTbasieq2}\\
&&{dq_{3}/d\tau}=i\Delta\Gamma q_{3}-ig^{\ast}q_{1}q_{2}, \label{TTbasieq3}
\end{eqnarray}
where some of the normalization to the parameters are borrowed from \cite{Phillips}:
\begin{eqnarray}
&&\tau=\eta z,\quad\eta={2d\over\pi c}\left({\omega_{1}\omega_{2}\omega_{3}\over n_{1}n_{2}n_{3}}\sum^{3}_{l=1}{n_{l}\over\omega_{l}}|\tilde{A}_{l0}|^{2}\right)^{1/2},\notag\\
&&q_{j}=A_{j}e^{i\left[\Delta k_{0}z-\int^{z}_{0}K_{\Lambda}(z')dz'\right]}/\left({\omega_{j}\over n_{j}}\sum^{3}_{l=1}{n_{l}\over\omega_{l}}|A_{l0}|^{2}\right)^{1/2},\notag\\
&&\Delta\Gamma=\left[\Delta k_{0}-K_{\Lambda}(\tau)\right]/\eta,\quad g=\Xi(\tau)e^{i\phi_{\mathrm{d}}(\tau)}. \label{characters}
\end{eqnarray}
Here, $z$ is the propagation distance; $n_{j}$, $\omega_{j}$, $A_{j}$ are the refractive index, angular frequency, and amplitude, respectively, of the slowly varying envelope of the $j$-th wave ($A_{j0}$ is $A_{j}$ at $\tau=0$); $d$ is the second-order nonlinear susceptibility; $\Delta k_{0}=k_{1}+k_{2}-k_{3}$ (where $k_{j}$ is the wave vector of the $j$-th wave) is the phase mismatch; and $K_{\Lambda}(\tau)$, $\Xi(\tau)=\sin[\pi D(\tau)]$ (with $0\leq D(\tau)\leq1$), and  $\phi_{\mathrm{d}}(\tau)$ are the wave vector, the effective duty cycle, and the phase of the QPM modulation, respectively. $|q_{j}|^{2}$ is proportional to the photon flux at the $j$-th wave, and their initial condition automatically satisfies $\sum^{3}_{j=1}|q^{(0)}_{j}|^{2}=1$ [cf. Eq. (\ref{Ainital}) in Appendix A]. The detailed derivation from the original slowly varying envelope equation of TWM to its dimensionless form is presented in Appendix A.

Eqs. (\ref{TTbasieq1}-\ref{TTbasieq3}) can be presented via a canonical Hamiltonian structure
\begin{eqnarray}
{dq_{j}\over d\tau}=-2i{\partial H \over \partial q^{\ast}_{j}}, \label{canonical}
\end{eqnarray}
where the Hamiltonian is $H={1\over2}\left(gq^{\ast}_{1}q^{\ast}_{2}q_{3}+g^{\ast}q_{1}q_{2}q^{\ast}_{3}\right)-{\Delta\Gamma\over2}\sum^{3}_{j=1}|q_{j}|^{2}$
The Manley-Rowe (MR) relations, which define 3 constants of motion for this Hamiltonian, are $K_{1}=|q_{1}|^{2}+|q_{3}|^{2}$, $K_{2}=|q_{1}|^{2}-|q_{2}|^{2}$, and $K_{3}=|q_{2}|^{2}+|q_{3}|^{2}$  \cite{Gil,Luther,Phillips}

\begin{figure}[t]
{\includegraphics[width=0.99\columnwidth]{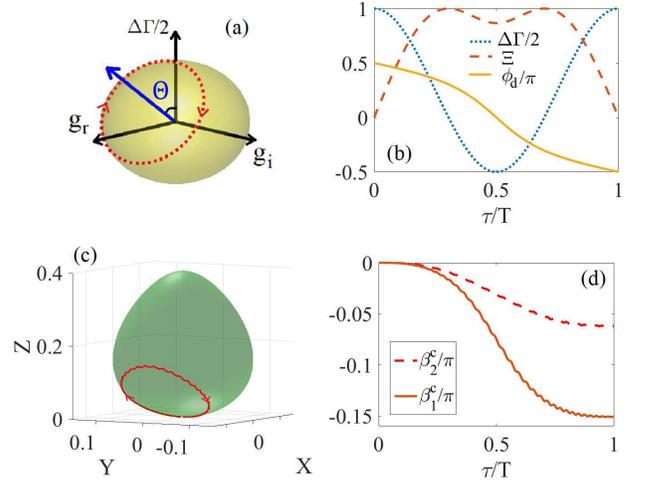}}
\caption{(a) A typical example of a clockwise circular rotation of the QPM vector $\mathbf{B}(\tau)$ on its parameter space sphere with $\Theta=\pi/4$. (b) The variations of the poling period $\Delta\Gamma$, the duty cycle $\Xi$ and phase $\phi_{\mathrm{d}}$, of the nonlinear coefficient along the nonlinear crystal in order to achieve $\Theta=\pi/4$. (c) A corresponding rotation of the state vector, $\mathbf{W}(\tau)$, which follows the clockwise rotation of the QPM vector in panel (a), draws a trajectory onto the state surface (\ref{Sphere}) with initial conditions $(I_{1},I_{2},I_{3})=(0.4,0.6,0)$ (where $I_{j}=|q^{0}_{j}|^{2}$). (d) AGP accumulations for $q_{1,2}$ in panel (c) along the scaled propagation coordinate.} \label{Vector}
\end{figure}

To show the geometric motion of Eqs. (\ref{TTbasieq1}-\ref{TTbasieq3}) in the state space, a set of coordinates $(X, Y, Z)$ and a surface $\varphi(X,Y,Z)=0$ are introduced by defining \cite{Luther,Phillips}
\begin{eqnarray}
&&X+iY=q_{1}q_{2}q^{\ast}_{3},\quad Z=|q_{3}|^2, \label{coordinate}\\
&&\varphi=X^2+Y^2-Z(Z-K_{1})(Z-K_{3}). \label{Sphere}
\end{eqnarray}
The shape and size (small or large) of this surface are defined by the initial condition of $q^{(0)}_{1,2,3}$. The Hamiltonian can be expressed by the coordinates (\ref{coordinate}) as
\begin{eqnarray}
H=\left(g_{\mathrm{r}}X+g_{\mathrm{i}}Y\right)+\Delta\Gamma[Z-(K_{1}+K_{3})]/2, \label{Ham2}
\end{eqnarray}
where $g_{\mathrm{r}}=\Xi\cos\phi_{\mathrm{\mathrm{d}}}$ and $g_{i}=\Xi\sin\phi_{\mathrm{d}}$ are the real and imaginary parts of $g$, respectively. Note that the imaginary part of $g$, which is created when $\phi_{\mathrm{d}}\neq0$ or $\pi$, can give rise to the $Y$ component of the Hamiltonian in the depleted case.

Based on the geometric definition in Eqs. (\ref{coordinate}-\ref{Ham2}), the dynamics of Eq. (\ref{canonical}) can be expressed in geometric form by defining a state vector $\mathbf{W}=X\hat{\mathbf{i}}+Y\hat{\mathbf{j}}+Z\hat{\mathbf{k}}$ and a QPM vector ${\mathbf{B}(\tau)=\nabla H=(g_{\mathrm{r}}, g_{\mathrm{i}},\Delta\Gamma/2)}$ as
\begin{eqnarray}
{d\mathbf{W}/d\tau}=\{\mathbf{W},H\}=\mathbf{B}(\tau)\times\nabla{\varphi}, \label{Weq}
\end{eqnarray}
where $\nabla=\hat{\mathbf{i}}\partial_{X}+\hat{\mathbf{j}}\partial_{Y}+\hat{\mathbf{k}}\partial_{Z}$. In the following, we will interpret the dynamics and assemble the AGP by means of this geometric description. For convenience, it is natural to fix the QPM vector as a unit vector (i.e., $|\mathbf{B}(\tau)|\equiv1$).

The typical processes for the TWM are the SFG and the difference-frequency generation (DFG). For the SFG process, the initial conditions are generally set as $q^{(0)}_{1,2}\neq0$ and $q^{(0)}_{3}=0$, which indicates that $\mathbf{W}$ is initiated at the south pole of the state surface. For the DFG process, we can set the initial conditions as $q^{(0)}_{2,3}\neq0$ and $q^{(0)}_{1}=0$. In this case, $\mathbf{W}$ is initiated at the north pole of the state surface. The adiabatic evolution requires to remain at an eigenstate over the entire process. Moreover, we require to complete the process at the initial state, so that the QPM vector must satisfy $\mathbf{B}(0)=\mathbf{B}(T)=\hat{\mathbf{k}}$. Hence, the circular rotation of the QPM vector can be expediently set as per \cite{Aviv},
\begin{eqnarray}
&&\Delta\Gamma(\tau)/2=\sin^{2}\Theta\cos\Omega\tau+\cos^{2}\Theta,\label{Detuning}\\
&&\Xi(\tau)=\sqrt{1-\left(\sin^{2}\Theta\cos\Omega\tau+\cos^{2}\Theta\right)^2},\label{Duty}\\
&&\phi_{d}(\tau)=\pm\arctan{\sin\Omega\tau\over\cos\Theta(1- \cos\Omega\tau)}, \label{phid}
\end{eqnarray}
where $\Theta$ is the angle of the normal vector of the plane of the circular trajectory [see Fig. \ref{Vector}(a)] and is defined with respect to the $\mathbf{\hat{k}}$ axis, $\Omega=2\pi/T$ is the angular velocity (here, $T=\eta L$ is the period of evolution, where $L$ is the length of the crystal). The positive and negative signs on the right-hand side of Eq. (\ref{phid}) refer to the different rotational directions. According to Eq. (\ref{characters}), the relationships between $\Theta$ and the other two QPM parameters are given by $K_{\Lambda}(\tau)=\Delta k_{0}-2\eta\left(\sin^{2}\Theta\cos\Omega\tau+\cos^{2}\Theta\right)$ and $D(\tau)=(1/\pi)\arcsin\sqrt{1-\left(\sin^{2}\Theta\cos\Omega\tau+\cos^{2}\Theta\right)^2}$. Since we assume a circular trajectory that starts at the upper top point in the parameter space [See in Fig. \ref{Vector}(a)], the angle $\Theta$ defines the trajectory in the parameter space and the corresponding variation of the state vector. A typical example of these QPM parameters varying along $\tau$ for a rotation is displayed in Fig. \ref{Vector}(b). This rotation can be imparted by binary modulation of the sign of $\chi^{(2)}$ along the nonlinear crystal, which can be realized for example by electric field poling of the ferroelectric crystals \cite{Reich,Hatanaka}. A typical example of a trajectory drawn by the rotation of the state vector $\mathbf{W}$, which follows the rotation of the QPM vector in Fig. \ref{Vector}(a), is displayed in Fig. \ref{Vector}(c).

\section{Phase difference between the clockwise and the counter-clockwise rotation}
\begin{figure}[tbp]
{\includegraphics[width=1.0\columnwidth]{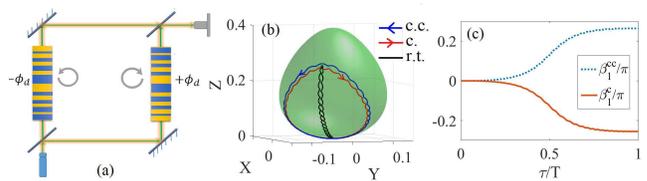}}%
\caption{(Color online) (a) Interferometric setup for measuring the phase difference between clockwise and counter-clockwise circular rotations. (b) Circular rotation for SFG. The blue and the red trajectories show the counter-clockwise (c.c.) and clockwise (c.) rotations, respectively. The black trajectory is the round-trip (r.t.) motion. Here, we select $\Theta=0.3\pi$ and $(I_{1},I_{2},I_{3})=(0.4,0.6,0)$. (c) Accumulation of $\beta_{1}$ for the clockwise (red solid curve) and counter-clockwise (blue dashed curve) in panel (b).}\label{Bottom}
\end{figure}

In the undepleted case, a Mach-Zhender interferometer can be used to measure the AGP  \cite{Wang2017,Aviv}, as shown in Fig. \ref{Bottom}(a), where in the two arms we have two crystals with opposite rotational direction, which are denoted respectively by $\pm\phi_{\mathrm{d}}$. The total phase shift in each of the two arms can be written as
\begin{eqnarray}
\pm\phi_{\mathrm{d}}:\quad\Phi^{\pm}_{j}=\mathfrak{D}_{j}\pm\beta_{j}, \label{totalphase}
\end{eqnarray}
where $\Phi^{\pm}_{j}$ is the total phase shift of $q_{j}$ in the different arms, $\mathfrak{D}_{j}$ represents the related dynamical phases, and $\beta_{j}$ is the accumulated geometric phase. Here, the definitions of the signs are adopted from Eq. (\ref{phid}). Based on Eq. (\ref{totalphase}), the magnitude of the geometric phase for wave $q_{j}$ can be calculated by
\begin{eqnarray}
\beta_{j}=(\Phi^{+}_{j}-\Phi^{-}_{j})/2=\Delta\Phi_{j}/2. \label{geometric}
\end{eqnarray}

We will extend this scheme to depleted case, and its validity will be verified by the analysis in the next section. The numerical simulations for the depleted SFG and the DFG process are conducted using Eqs. (\ref{TTbasieq1}-\ref{TTbasieq3}) with two different types of initial conditions, which are initiated from the two poles of the state surface. The algorithm for the numerical simulation is the 4-step Runge-Kutta method \cite{RK4}. In the simulation, we select the length of the crystal as $L=T/\eta=5$ cm with $\eta=25$ cm$^{-1}$, hence, $T=125$,  which is the period of the rotation and sufficient in practice for an adiabatic evolution \cite{Suchowski}.


\begin{figure}[tbp]
{\includegraphics[width=1\columnwidth]{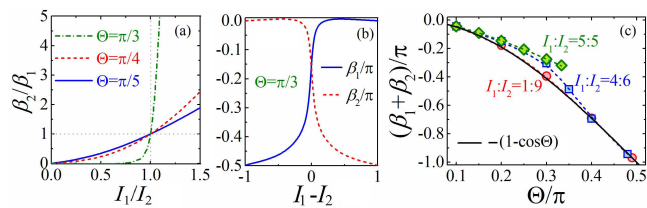}}%
\caption{(Color online) Circular rotation for SFG. (a) $\beta_{2}/\beta_{1}$ versus $I_{1}/I_{2}$ with different values of $\Theta$. (b)$\beta_{1,2}$ versus $I_{1}-I_{2}$ for $\Theta=\pi/3$. (c) The sum of AGP for $q_{1,2}$ versus $\Theta$ with different depletion level. Here, the symbols (circle, square, and rhombus) represent the numerical results from the interferometric scheme in Eq. (\ref{geometric}), while the dashed curve is the prediction from Eqs. (\ref{geometric3}) and (\ref{geometric4}). The black curve is the theoretical result from Eq. (\ref{undepletedAGP}) for the undepleted case.}\label{beta1beta2}
\end{figure}

For the SFG process, we assume that the initial conditions satisfy $I_{2}\geq I_{1}>0$ and $I_{3}=0$ (note that $I_{j}=|q^{(0)}_{j}|^{2}$). If $q^{(0)}_{1,2}$ satisfy $I_{1}\ll I_{2}$, the system returns to the undepleted case, which is initiated by a weak idler ($q_{1}$) wave and an undepleted strong pump ($q_{2}$) wave. If these waves satisfy $I_{1}=I_{2}$, the system undergoes a fully depleted SFG process. Hence, the system's degree  of depletion can be controlled by the ratio $I_{1}/I_{2}$.

With these initial conditions, we find that the state vector $\mathbf{W}(\tau)$ starts from the south pole of the state surface (\ref{Sphere}) and draws a circular trajectory following the slow rotation of the QPM vector $\mathbf{B}(\tau)$ [see an example for a clockwise and a counterclockwise circular rotation in Fig. \ref{Bottom}(b)]. The positive and negative signs in Eq. (\ref{phid}) refer to the clockwise and the counter-clockwise rotation, respectively, which are shown by the red and blue curves in Fig. \ref{Bottom}(b). Because $q_{3}(0)=q_{3}(T)=0$ and the AGPs are accumulated in the original frequencies, only the AGPs of $q_{1,2}$ are taken into account for this case. A typical example of the accumulations of $\beta_{1,2}$ (for a clockwise circular rotation) are displayed in Fig. \ref{Vector}(d).

The ratio of  $\beta_{2}/\beta_{1}$ as a function of $I_{1}/I_{2}$ is displayed in Fig. \ref{beta1beta2}(a). This panel shows that the AGP of $q_{1}$ is the dominant component when $I_1 < I_2$ , i.e. the phase accumulation in $q_1$ is much larger than in $q_2$. The dominant role of $q_1$ can be further enhanced by increasing the value of $\Theta$. It is interesting to note that small change in the ratio $I_{1}/I_{2}$ near the point in which they are equal, e.g. at the intersection point of $(I_{1}/I_{2},\beta_{2}/\beta_{1})=(1,1)$ and for the case $\theta=\pi/3$, leads to dramatic change of the ratio $\beta_{2}/\beta_{1}$. Note that this calculation is still valid if we exchange $I_1$ and $I_2$, which means that the large phase modulation is now accumulated at $\omega_{2}$ (rather than at $\omega_{1}$), when $I_{1}>I_{2}$. All of these relations are analytically predicted by Eq. (\ref{geometric4}) in Section 4. Fig. \ref{beta1beta2}(b) displays the detailed variations of $\beta_{1,2}$ as a function of the difference of $I_{1}-I_{2}$.

Fig. \ref{beta1beta2}(c) displays the sum of the two phases, $\beta_{1}+\beta_{2}$, as a function of $\Theta$ for different ratios of $I_{1}$ and $I_{2}$. When the depletion is weak, i.e., $I_{1}\ll I_{2}$, $\beta_{2}\approx0$, and $\beta_{1}$ is very close to the analytical prediction of the AGP in the undepleted case \cite{Aviv}
\begin{eqnarray}
\beta_{1}=-\pi(1-\cos\Theta). \label{undepletedAGP}
\end{eqnarray}
Eq. (\ref{undepletedAGP}) implies the clockwise rotation acquires a negative value of AGP. If the rotation direction switches to a counter-clockwise type, a positive value of AGP is acquired. Therefore, the AGP in the process of SFG satisfies a right-hand chirality. Left-hand chirality would be obtained by performing a DFG process.

\section{Analytical description}

The numerical results we obtained in the previous section can be verified by the following theoretical analysis: under the adiabatic condition, the variation of states $q_{j}$ in the whole process remains an eigenstate with a geometric phase factor as
\begin{eqnarray}
q_{j}(\tau)=\tilde{q}_{j}e^{i\beta_{j}(\tau)}, \label{splitted}
\end{eqnarray}
where $\tilde{q}_{j}$ and $\beta_{j}$ are the eigenstate and the accumulated AGP for the $j$-th wave, respectively. Here, we provide the derivation of $\beta_{j}$ by using $j=1$ as an example. Substituting Eq. (\ref{splitted}) with $j=1$ into Eq. (\ref{TTbasieq1}),
\begin{eqnarray}
i\mu_{1}\tilde{q}_{1}+i\tilde{q}_{1}\left(d\beta_{1}/d\tau\right)=i\Delta\Gamma\tilde{q}_{1}-igq^{\ast}_{2}q_{3}e^{-i\beta_{1}}, \label{dd2}
\end{eqnarray}
where $\mu_{1}$ is the eigenvalue of $\tilde{q}_{1}$ because $d\tilde{q}_{1}/d\tau=i\mu_{1}\tilde{q}_{1}$. Multiply Eq. (\ref{dd2}) by $\tilde{q}^{\ast}_{1}$, and noting that $|q_{1}|^{2}=|\tilde{q}_{1}|^{2}$ and $\tilde{q}^{\ast}_{1}=q_{1}e^{i\beta_{1}}$; this yields $d\beta_{1}/d\tau=\left(\Delta\Gamma|q_{1}|^{2}-gq^{\ast}_{1}q^{\ast}_{2}q_{3}\right)/|q_{1}|^{2}-\mu_{1}$. Therefore, the evolution of $\beta_{1}$ can be obtained by
\begin{eqnarray}
\beta_{1}(\tau)=\int^{\tau}_{0}\left[\left(\Delta\Gamma|q_{1}|^{2}-gq^{\ast}_{1}q^{\ast}_{2}q_{3}\right)/|q_{1}|^{2}-\mu_{1}\right]d\tau'. \label{geometric2}
\end{eqnarray}

By applying Eq. (\ref{canonical}), the overall AGP of $\beta_{1}$ can also be obtained by rewriting Eq. (\ref{geometric2}) as
\begin{eqnarray}
\beta_{1}=\int^{T}_{0}\left(-\frac{1}{|q_{1}|^{2}}2q^{\ast}_{1}{\partial H\over\partial q^{\ast}_{1}}\right)d\tau-\int^{T}_{0}\mu_{1} d\tau\notag=\Phi_{1}-\mathfrak{D}_{1},\notag\\ \label{geometric3}
\end{eqnarray}
where $\Phi_{1}$ and $\mathfrak{D}_{1}$ are the total phase differences and the dynamical phase of $q_{1}$, respectively.

For the fully depleted case, i.e., the case of $I_{1}=I_{2}$, eigenvalues $\mu_{1,2}$ have an explicit form as \cite{Gil}
\begin{eqnarray}
\mu_{1,2}=\left[5\Delta\Gamma+\sqrt{\Delta\Gamma^{2}+5|g|^{2}(K_{1}+K_{3})}\right]/6, \label{eigenvalue}
\end{eqnarray}
where $K_{1,3}$ are given by the MR relation. The top dashed curve and the rhombuses in Fig. \ref{beta1beta2}(c) represent the comparison between $\beta_{1,2}$ and Eqs. (\ref{geometric2}) and  (\ref{geometric}) for circular rotation with different values of $\Theta$. The agreement between these two results demonstrates the validity of the analytical prediction for the fully depleted cases. If the rotation is switched to the counter-clockwise type, i.e., $g=\Xi e^{-i\phi_{\mathrm{d}}}$, the results satisfy $\beta^{\mathrm{cc}}_{j}=-\beta^{\mathrm{c}}_{j}>0$ [see an example in Fig. \ref{Bottom}(c)], which clearly indicates that the interferometric scheme in section III remains valid for the fully depleted cases. Because this scheme was verified by two limits of depletion (undepleted case and fully depleted case), one can expect that this scheme remains valid for any level of depletion. In the Section II or the Appendix B, we will explain why the green dashed curve (i.e., $I_{1}=I_{2}$) stops at $\Theta=0.35\pi$.

The eigenvalues $\mu_{j}$ are independent of $\phi_{\mathrm{d}}$. This was shown analytically in Eq. (\ref{eigenvalue}) for the case of the full depletion, and was confirmed by us numerically for other levels of depletion. The motion of eigenstates, which is governed by the QPM vector defined as per Eqs. (\ref{Detuning}) and (\ref{Duty}), creates a round-trip trajectory on the state surface [see the black curve in Fig. \ref{Bottom}(b)]. By extending the subscript from $j=1$ to 3 and applying a following transformation $q^{\prime}_{j}=q_{j}/N_{j}$, $p^{\prime}_{j}=-iq^{\ast}_{j}/N_{j}$ (where $N_{j}=\sqrt{2|q_{j}|^{2}}$), Eq. (\ref{geometric3}) can be further adjusted to a more concise form as
\begin{eqnarray}
\beta_{j}=\int_{\bigcirc} p^{\prime}_{j}dq^{\prime}_{j}-\int_{\upharpoonleft\downharpoonright}p^{\prime}_{j}d q^{\prime}_{j}=\Phi^{c}_{j}-\Phi^{\mathrm{rt}}_{j},\label{geometric4}
\end{eqnarray}
where $\Phi^{c}_{j}$, and $\Phi^{\mathrm{rt}}_{j}$ stand for the total phase of the $j$-th wave after circular (clockwise or counter-clockwise) rotation and round-trip motion, respectively. The subscripts `$\bigcirc=\circlearrowright$ or $\circlearrowleft$' denote the clockwise or the counter-clockwise rotations and `$\upharpoonleft\downharpoonright$' denotes round-trip motion. Eq. (\ref{geometric4}) implies that the acquired AGP for the entire process can be viewed as the difference of the action-angle variables in the phase space of $(q^{\prime}_{j}, p^{\prime}_{i})$ \cite{Jieliu} between two kinds of motion (the circular rotation and round-trip motion).

According to the above discussions, these three kinds of motions (clockwise, counter-clockwise circular rotations and round-trip motion) share the same $\mu_{j}$. orrespondingly, numerical simulations reveal that the Hamiltonian of each of these motions obtains the same value at each point through the entire process. In spite of this degeneracy, the phases that the waves accumulate in each of this motions is different. We found numerically that the total phase of the round-trip motion is $\Phi^{\mathrm{rt}}_{j}=\int\mu_{j}d\tau'\equiv \mathfrak{D}_{j}$, and the total phases of the clockwise and the counter-clockwise are $\Phi^{c}_{j}=\mathfrak{D}_{j}\pm|\beta^{c}_{j}|$, respectively, which is exactly the same as Eq. (\ref{totalphase}). Hence, the consistence between Eq. (\ref{totalphase}) and Eq. (\ref{geometric4}), which is also demonstrated by the numerical simulations, indicates that the interferometric scheme is valid for all levels of depletion.

Because the circular rotation is manipulated by $(\Delta\Gamma,\Xi,\phi_{\mathrm{d}})$, while the round-trip motion is driven by only $(\Delta\Gamma,\Xi)$, the trajectory of $\mathbf{W}(\tau)$ of round-trip motion can be viewed as the vertical projection of its counterparts in the circular and counter-circular rotations [See in Fig. \ref{Bottom}(b)]. Hence, the AGP can be also viewed as the difference between the total phase accumulated along the circular trajectory with respect to its vertical projection in the space of $(X,Y,Z)$. A visualization for these three types of motions with $(I_1, I_2 ) = (0.4, 0.6)$ and $\Theta = 0.3\pi$ are shown by a movie in the Supplemental Material \cite{RK4}.


Eq. (\ref{geometric4}) is universal for the undepleted (linear) and the depleted (nonlinear) cases. It can also simplify the calculation of AGP for any degrees of depletion. The results in Fig. \ref{beta1beta2}(c) (the dashed curves) demonstrate that Eq. (\ref{geometric4}) for $\beta_{1,2}$ can match the interferometric scheme in Section 3 perfectly for all levels of depletion.

Finally, it is necessary to point out that the AGPs in Eq. (\ref{geometric}) satisfy $|\beta_{j}|\sim N^{-2}_{j}\sim I^{-1}_{j}$, which explains the reason why $\beta_{2}/\beta_{1}<1$ with $I_{1}<I_{2}$ or vice versa in Section 3.

\begin{figure}[tbp]
{\includegraphics[width=1\columnwidth]{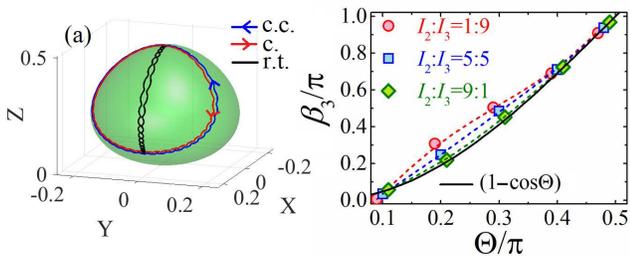}}%
\caption{(Color online) Circular rotation in the case of DFG. (a) The blue and red trajectories are the counter-clockwise (c.c.) and clockwise (c.) rotations, respectively, with $\Theta=\pi/3$ for initial conditions $(I_{1},I_{2},I_{3})=(0,0.5,0.5)$. The black curve represents the round-trip case. (b) The AGP for $q_{3}$ (i.e., $\beta_{3}$) as a function of $\Theta$ for different ratios $I_{2}/I_{3}$. The symbols were calculated with Eq. (\ref{geometric}) The dashed lines were calculated with Eq. (\ref{geometric4}). The black solid curve represents the undeplete case.}\label{Top}
\end{figure}

\section{AGP in the depleted DFG process}
For the DFG process, we assume that the initial conditions are $I_{2,3}>0$ and $I_{1}=0$. The rotation in DFG starts from the north pole of the state surface (i.e., $Z=|q^{(0)}_{3}|^{2}$). In this case, the positive and negative signs in Eq. (\ref{phid}) remain to refer to the clockwise and the counter-clockwise rotation, respectively. Because $q_{1}(0)=q_{1}(T)=0$ and the AGPs are accumulated in the original frequencies, we consider only the phase information for $q_{2}$ and $q_{3}$ in the following discussion.

Fig. \ref{Top}(a) shows typical examples of a counter-clockwise (blue curve) and a clockwise (red curve) circular rotation with $I_{2}=I_{3}=1/2$ and $\Theta=\pi/3$. In contrary to the case of SFG, numerical simulations show that $\beta_{3}$ plays a dominant role regardless of wether $I_{2}>I_{3}$ or $I_{2}<I_{3}$. Hence, the geometric phase is accumulated mainly in $q_{3}$ in this case. Fig. \ref{Top}(b) displays the AGP, which is represented by $\beta_{3}$ (we ignore the contribution from $\beta_{2}$ here) as a function of $\Theta$ with different depleted conditions. Unlike the case of SFG, the AGPs in the depleted DFG are all positive. According to Eq. (\ref{geometric}), the clockwise rotation gives a positive value, which means that the AGP has a left-hand chirality. Hence, the SFG and the DFG process have a different chirality for acquiring the AGP. Moreover, the numerical simulations demonstrate that Eq. (\ref{geometric4}) with $j=3$ is valid in predicting the AGP for the DFG process (see the different dashed curves in Fig. \ref{Top}b).

\section{Application}

\begin{figure}[tbp]
{\includegraphics[width=0.9\columnwidth]{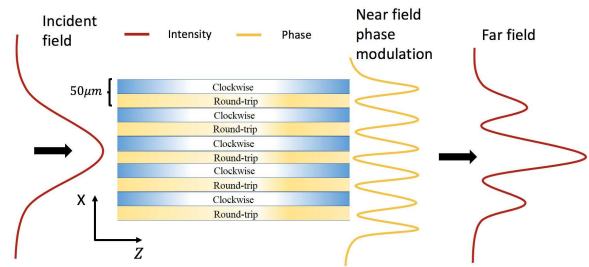}}%
\caption{(Color online) Schematic drawing of the grating simulation. Each segment of poling in the $x$ direction corresponds to clockwise rotation and round-trip motion in parameter space. The period of each segment is set to be $\Lambda=50$ $\mu$m. }\label{AGPgrating}
\end{figure}

\begin{figure}[tbp]
{\includegraphics[width=1\columnwidth]{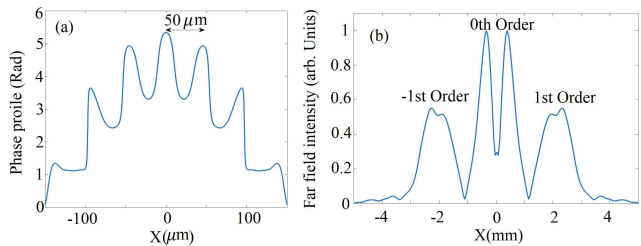}}%
\caption{(Color online)(a)Phase profile of the beam at the end of the geometric phase grating. (b) Far field intensity after the crystal end. }\label{PGandFarFI}
\end{figure}
The control of the adiabatic geometric phase opens exciting new possibilities for all-optical manipulation of light. The first potential application is a power-control phase modulator. In Figs. \ref{beta1beta2}(a,b), the strong dependence of the phase on the intensities may enable interesting new possibilities for all-optical phase modulation of one optical wave by another optical wave. For example, small variations in the intensity of $I_1$ near the equal intensities point will induce significant variations in the phase of $\beta_2$.

The other potential application is the AGP grating, which can be utilized for splitting the beam into different diffraction orders in the far field. Unlike the case of SFG with undepleted pump, here we consider the fully nonlinear second harmonic generation process, hence only a single input wave is needed. The crystal, which is assumed to be LiNbO$_{3}$, is modulated by stripes of alternative circular rotation and round-trip motion in the direction perpendicular to the propagation of the light (See in Fig. \ref{AGPgrating}). The poling is switched between clockwise rotation and round-trip motion in parameter space with $\Theta=\pi/4$ and a period of $\Lambda=50$ $\mu$m (See in Fig. \ref{AGPgrating}). The crystal has the following attributes: $d_{33}=23$ pm/V, $L_{z}=1$ cm, $L_{x}$=2 mm. The laser used in the simulation has a beam waist of 200 $\mu$m, peak power of 350 kW, and $\lambda_{\mathrm{FH}}=1.064$ $\mu$m (where the subscript `FH' means the the first harmonic). The simulation assumes a planar waveguide configuration, where the diffraction in the $y$ axis can be neglected.

The angular distance between diffraction orders is given by:
\begin{eqnarray}
X={\lambda_{Z}\over\Lambda}=0.0021\mathbf{m} \label{angulardistance}
\end{eqnarray}
where $f=0.1$ m is the focal distance of the lens used for imaging the crystal output. This angular distance along with the phase profile at the crystal end can be seen in Fig. \ref{PGandFarFI}(a), and the far field intensity after the crystal end is shown in Fig. \ref{PGandFarFI}(b).

\section{Conclusion}

In conclusion, we proposed a scheme of circular rotation for accumulating AGP in the fully nonlinear regime of three wave mixing and analyzed the accumulated phase for the three interacting waves. This adiabatic geometric phase is controlled by a slow variation in the three quasi-phase-matching (QPM) parameters, the period $K_{\Lambda}(z)$, the duty cycle $D(z)$, and the phase $\phi_{\mathrm{d}}(z)$. These three QPM parameters define a vector in the parameter space and carry out a circular rotation. Under the adiabatic condition, the state vector follows the rotation of the QPM vector and draws a circular trajectory on the state surface, and the AGP can be evaluated by an interferometric scheme and predicted by a theoretical analysis. The theoretical analysis indicates that the AGP equals the difference between the accumulated phase along the circular trajectory (for either clockwise or counter-clockwise) and the round-trip trajectory, which can be viewed as the vertical projection of the same circular trajectory. Our analysis is in an excellent agreement with the undepleted case   when the intensity of one of the two input waves is much larger than the other. The different chiralities for the sum-frequency generation and the difference-frequency generation are demonstrated throughout our analysis. Our findings open new possibilities for all-optical manipulation of light, without the need of a strong and nearly undepleted pump. This is specifically relevant for efficient second harmonic generation process, in which the pump must be depleted to enable efficient conversion. Moreover, we have identified conditions in which the accumulated phase strongly depends on the intensities of the interacting waves, thus enabling all-optical modulation of the phase of one wave by the intensity of the other wave.

\begin{acknowledgments}
 This work was supported by the Israel Science Foundation (ISF), grant no. 1415/17, by the NNSFC (China) through Grant No. 11874112, 11974146, \& 11575063, by the Natural Science Foundation of Guangdong Province (China) through Grant No. 2017B030306009, and by the State Scholarship Fund of China Scholarship council through File No. 201808440001.
\end{acknowledgments}


\newpage

\appendix
\section{Scaled form of the dynamical coupled wave equations for the three-wave mixing}
The dynamical equations of three-wave mixing with slowly varying envelope are given by
\begin{eqnarray}
&&\partial_{z}A_{1}=-i{2d(z)\omega_{1}\over c n_{1}}A^{\ast}_{2}A_{3}e^{-i\Delta k_0 z}\label{Basieq1}\\
&&\partial_{z}A_{2}=-i{2d(z)\omega_{2}\over c n_{2}}A^{\ast}_{1}A_{3}e^{-i\Delta k_0 z}\label{Basieq2}\\
&&\partial_{z}A_{3}=-i{2d(z)\omega_{3}\over c n_{3}}A_{1}A_{2}e^{i\Delta k_0 z}, \label{Basieq3}
\end{eqnarray}
where $A_{1,2,3}$ are the slowly varying envelopes of the idler, pump, and signal waves, respectively; $\Delta k_{0}=k_{1}+k_{2}-k_{3}$ is the phase mismatch; $d(z)$ is the spatially varying magnitude of the second-order nonlinear susceptibility, which can be described by the Fourier series with slowly varying components as
\begin{eqnarray}
&&d(z)=d_{ij}\sum^{\infty}_{m=-\infty}|d_{m}(z)|\times\notag\\
&&\exp\left\{im\left[\int^{z}_{0}K_{\Lambda}(z')dz'+\phi_{\mathrm{d}}(z)\right]\right\}, \label{Dz1}
\end{eqnarray}
with
\begin{eqnarray}
d_{m}(z)=
\begin{cases}
(2/ m\pi)\sin\left[m\pi D(z)\right] & m\neq0 \\
2D(z)-1 & m=0
\end{cases}.\label{Dz2}
\end{eqnarray}
The detailed meaning of the QPM (quasi-phase-matching) parameters in Eqs. (\ref{Dz1},\ref{Dz2}) are available in TABLE I.

\begin{table}[!h]
\caption{\label{tableI}The detailed meaning of the characters in Eqs. (\ref{Dz1},\ref{Dz2}) }
\begin{tabular}{lcr}
\hline\hline
\specialrule{0em}{1pt}{1pt}
$K_{\Lambda}(z)$ &$\quad$ & modulation wave-vector \\
\specialrule{0em}{1pt}{1pt}
$\phi_{d}(z)$ &$\quad$ & modulation phase \\
\specialrule{0em}{1pt}{1pt}
$0\leq D(z)\leq1$ &$\quad$ & duty cycle \\
\specialrule{0em}{1pt}{1pt}
$d_{ij}$ & $\quad$ & nonlinear susceptibility tensor\\
\specialrule{0em}{1pt}{1pt}
\hline
\end{tabular}
\end{table}

If we only select $m=\pm1$ from Eq. (\ref{Dz1}), which are assumed to provide best compensation for the phase mismatch, equation (\ref{Dz1}) can be simplified to
\begin{eqnarray}
&&d(z)=d_{ij}\left(2\over\pi\right)\sin\left(\pi D(z)\right)\times\notag\\
&&\left\{e^{+i\left[\int^{z}_{0}K_{\Lambda}(z')dz'+\phi_{d}(z)\right]} +e^{-i\left[\int^{z}_{0}K_{\Lambda}(z')dz'+\phi_{d}(z)\right]}\right\}.\notag\\ \label{PNf}
\end{eqnarray}
The positive and negative frequency component of Eq. (\ref{PNf}) can be substituted into Eqs. (\ref{Basieq1},\ref{Basieq2}) and Eq. (\ref{Basieq3}), respectively (which is similar to the application of rotating wave approximation) \cite{Aviv}.
Making a transformation of $A_{1,2,3}$ to a rotating frame $\tilde{A}_{1,2,3}$ by
\begin{eqnarray}
&&A_{1}=\tilde{A}_{1}\exp\left\{-i\left[\Delta k_{0}z-\int^{z}_{0}K_{\Lambda}(z')dz'\right]\right\},\notag\\
&&A_{2}=\tilde{A}_{2}\exp\left\{-i\left[\Delta k_{0}z-\int^{z}_{0}K_{\Lambda}(z')dz'\right]\right\},\notag\\
&&A_{3}=\tilde{A}_{3}\exp\left\{-i\left[\Delta k_{0}z-\int^{z}_{0}K_{\Lambda}(z')dz'\right]\right\}. \label{RFT}
\end{eqnarray}
and substituting them into Eqs. (\ref{Basieq1},\ref{Basieq2} \ref{Basieq3}), the three-wave equations are changed to
\begin{eqnarray}
&&\partial_{z}\tilde{A}_{1}=i\left[\Delta k_{0}-K_{\Lambda}(z)\right]\tilde{A}_{1}\notag\\
&&-i{2(\pi/2)d_{ij}\omega_{1}\over c n_{1}}\sin\left[\pi D(z)\right]e^{i\phi_{d}(z)}\tilde{A}^{\ast}_{2}\tilde{A}_{3}\label{Tbasieq1}\\
&&\partial_{z}\tilde{A}_{2}=i\left[\Delta k_{0}-K_{\Lambda}(z)\right]\tilde{A}_{2} \notag\\
&&-i{2(\pi/2)d_{ij}\omega_{2}\over c n_{2}}\sin\left[\pi D(z)\right]e^{i\phi_{d}(z)}\tilde{A}^{\ast}_{1}\tilde{A}_{3}\label{Tbasieq2}\\
&&\partial_{z}\tilde{A}_{3}=i\left[\Delta k_{0}-K_{\Lambda}(z)\right]\tilde{A}_{3} \notag\\
&&-i{2(\pi/2)d_{ij}\omega_{3}\over c n_{3}}\sin\left[\pi D(z)\right]e^{-i\phi_{d}(z)}\tilde{A}_{1}\tilde{A}_{2}\label{Tbasieq3}
\end{eqnarray}
By applying following definitions \cite{Phillips}:
\begin{eqnarray}
&&\tilde{A}_{j}=q_{j}\left({\omega_{j}\over n_{j}}\sum^{3}_{l=1}{n_{l}\over\omega_{l}}|\tilde{A}_{l0}|^{2}\right)^{1/2},\notag\\
&&\eta={2d_{ij}\over\pi c}\left({\omega_{1}\omega_{2}\omega_{3}\over n_{1}n_{2}n_{3}}\sum^{3}_{l=1}{n_{l}\over\omega_{l}}|\tilde{A}_{l0}|^{2}\right)^{1/2},\notag\\
&&\tau=\eta z,\notag\\
&&\Delta\Gamma(\tau)=\left[\Delta k_{0}-K_{\Lambda}(\tau)\right]/\eta,\notag\\
&&g(\tau)=\sin\left[\pi D(\tau)\right]e^{i\phi_{d}(\tau)}=\Xi(\tau)e^{i\phi_{d}(\tau)}, \label{Acharacters}
\end{eqnarray}
equations (\ref{Tbasieq1}-\ref{Tbasieq3}) can be changed to
\begin{eqnarray}
&&{dq_{1}\over d\tau}=i\Delta\Gamma q_{1}-ig(\tau)q^{\ast}_{2}q_{3}, \label{ATTbasieq1}\\
&&{dq_{2}\over d \tau}=i\Delta\Gamma q_{2}-ig(\tau)q^{\ast}_{1}q_{3}, \label{ATTbasieq2}\\
&&{dq_{3}\over d \tau}=i\Delta\Gamma q_{3}-ig(\tau)^{\ast}q_{1}q_{2}. \label{ATTbasieq3}
\end{eqnarray}

Obviously, according to the definitions in Eq. (\ref{Acharacters}), the initial condition for Eqs. (\ref{ATTbasieq1}-\ref{ATTbasieq3}) satisfies
\begin{eqnarray}
&&\sum^{3}_{j}|q^{(0)}_{j}|^2=\sum_{j=1}^{3}{|A_{j0}|^{2}\over\left({\omega_{j}\over n_{j}}\sum_{l=1}^{3}{n_{l}\over\omega_{l}}|A_{l0}|^{2}\right)}\notag\\
&&={1\over\sum_{l=1}^{3}{n_{l}\over\omega_{l}}|A_{l0}|^{2}}\sum_{j=1}^{3}{n_{j}\over\omega_{j}}|A_{j0}|^{2}\equiv1. \label{Ainital}
\end{eqnarray}

\section{Cutoff of AGP curve at $\Theta=0.35\pi$ for $I_{1}=I_{2}$}
\begin{figure}[tbp]
{\includegraphics[width=0.9\columnwidth]{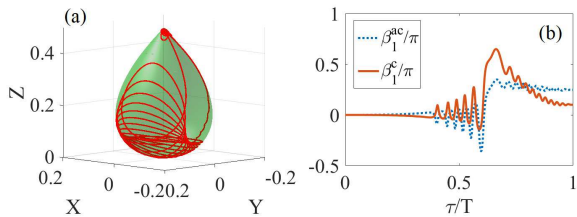}}%
\caption{(Color online) (a) Trajectory of the $\mathbf{W}(\tau)$ of clockwise rotation with $\Theta=0.4\pi$ in the full depleted case (i.e. $I_{1}=I_{2}$). (b)The evolutions of the phase of $\beta_{1}$ for the clockwise and counter-clockwise rotation, which are calculated via Eq. (\ref{geometric4}). }\label{Thetath}
\end{figure}

At the case of $I_{1}=I_{2}$, the system is under the fully depleted condition. At this case, Ref. \cite{Gil} had proved that the eigenstate $\tilde{q}_{1,2}\equiv0$ at $\Delta\Gamma<\Delta\Gamma_{\min}$, where
\begin{eqnarray}
\Delta\Gamma_{\min}=-\sqrt{2(K_{1}+K_{3})}=-\sqrt{2}. \label{Gammamin}
\end{eqnarray}
Here $K_{1,3}$ are two constants which are given by the MR relations. According to Eq. (\ref{Ainital}), $K_{1}+K_{3}=1$.

According to the definition in Eq. (\ref{Detuning}), one may obtain $\Theta_{\max}\approx0.4\pi$ at $\Delta\Gamma=\Delta\Gamma_{\min}$. This result implies that the curve of $\beta_{1,2}(\Theta)$ must be stop at $\Theta=\Theta_{\max}$ in the limit of the full depletion.

Actually, Eq. (\ref{Gammamin}) did not consider the circular rotation of the QPM vector. It was derived by fixing $\Xi(\tau)\equiv1$. Hence, for the case of circular rotation,  i.e., $\Xi(\tau)\in[0,1]$, the value of $\Theta_{\mathrm{\max}}$ needs to be adjusted. The numerical simulations demonstrated that the curve of $\beta_{1,2}(\Theta)$ stops at $\Theta_{\mathrm{\max}}\approx0.35\pi$ for the circular rotation [see the top dashed curve and rhombus in Fig. \ref{beta1beta2}(c)].

At the limit of $I_{1}=I_{2}$, because the eigenstates $\tilde{q}_{1,2}=0$ in the case of $\Theta>\Theta_{\max}$. Under this circumstance, the state vector $\mathbf{W}$ can reach the north pole of the surface. A typical example of the trajectory of $\mathbf{W}(\tau)$ for this case is displayed in Fig. \ref{Thetath}(a), which shows that a strong helical trajectory is produced after the $\mathbf{W}$ passing through the north pole. Moreover, in Ref. \cite{Gil}, at the cusp, the eigenstates cannot follow the equivalent magnetic field vector since the adiabatic conditions are no longer satisfied for the parameters of this calculation. Fig. \ref{Thetath}(b) shows that the process of accumulating the AGP in the case of Fig. \ref{Thetath}(a), which manifests the failure in the generation of the AGP.

\end{document}